\documentclass[a4paper,aps,prl,twocolumn,a4paper]{revtex4}
\usepackage{epsfig}
\usepackage{graphicx}
\usepackage{epsf}
\usepackage{bm}
\usepackage[T1]{fontenc}
\usepackage{amsmath,amssymb}
\usepackage[latin1]{inputenc}
\usepackage{times}
\usepackage{ulem}
\usepackage{color}
\usepackage[squaren,Gray]{SIunits}

\usepackage[squaren,Gray]{SIunits}
\usepackage{times}

\usepackage{amsmath}
\usepackage{amsfonts}
\usepackage{amssymb}
\usepackage{graphicx}
\usepackage[latin1]{inputenc}
\usepackage{color}
\newcommand{\cora}[1]{\textcolor{black}{#1}}
\newcommand{\coraa}[1]{\textcolor{black}{#1}}

\begin{document}

\title{Smart optical coherence tomography for ultra-deep imaging through highly scattering media}

\author{Amaury Badon}
\author{Dayan Li}
\author{Geoffroy Lerosey}
\author{A. Claude Boccara}
\author{Mathias Fink}
\author{Alexandre Aubry}
\email{alexandre.aubry@espci.fr}
\affiliation{Institut Langevin, ESPCI Paris, PSL Research University, CNRS UMR 7587, 1 rue Jussieu, F-75005 Paris, France}

\date{\today}

\begin{abstract}
Multiple scattering of waves in disordered media is a nightmare whether it be for detection or imaging purposes. The best approach so far to get rid of multiple scattering is optical coherence tomography. It basically combines confocal microscopy and coherence time-gating to discriminate ballistic photons from a predominant multiple scattering background. Nevertheless, the imaging depth range remains limited to 1 mm at best in human soft tissues. Here we propose a matrix approach of optical imaging to push back this fundamental limit. By combining a matrix discrimination of ballistic waves and iterative time-reversal, we show both theoretically and experimentally an extension of the imaging-depth limit by at least a factor two compared to optical coherence tomography. In particular, the reported experiment demonstrates imaging through a strongly scattering layer from which only one reflected photon over 1000 billion is ballistic. This approach opens a new route towards ultra-deep tissue imaging.
\end{abstract}

\maketitle

\section*{INTRODUCTION}

The propagation of light in inhomogeneous media is a fundamental problem with important applications, ranging from astronomical observations through a turbulent atmosphere to deep tissue imaging in microscopy or light detection through a dense cloud in LIDAR technology. Conventional focusing and imaging techniques based on the Born approximation generally fail in strongly scattering media due to the multiple scattering (MS) events undergone by the incident wavefront. Recent advances in light manipulation techniques have allowed great progresses in optical focusing through complex media \cite{Mosk_review}. Following pioneering works in ultrasound \cite{derode0,derode_2003}, Vellekoop and Mosk \cite{Vellekoop} showed how light can be focused spatially through a strongly scattering medium by actively shaping the incident wavefront with a spatial light modulator (SLM). Subsequently, a matrix approach of light propagation through complex media was developed \cite{popoff}. It relies on the measurement of the Green's functions between each pixel of a SLM and of a charge-coupled device (CCD) camera across a scattering medium. The experimental access to this transmission matrix allows to take advantage of MS for optimal light focusing \cite{popoff,choi,salma} and communication \cite{popoff2,Cizmar} across a diffusive layer or a multi-mode fiber. 

MS is a much more difficult challenge with regards to imaging. On the one hand, imaging techniques like diffuse optical tomography \cite{Durduran}, acousto-optic \cite{Resink} or photoacoustic \cite{Xu} imaging take advantage of the diffuse light to image scattering media in depth but their resolution power is limited. More recently, the memory-effect exhibited by the MS speckle \cite{freund,feng} has been taken advantage of to image objects through strongly scattering layers with a diffraction-limited resolution \cite{bertolotti,katz_2012,katz_2014}. However, it only applies to thin opaque layers as the field-of-view is inversely proportional to the scattering medium thickness. On the other hand, conventional reflection imaging methods provide an optical diffraction-limited resolution but usually rely on a single scattering (SS) assumption. The imaging limit of conventional microscopy can be derived from the scattering mean free path $l_s$. It describes the average distance that a photon travels between two consecutive scattering events. In turbid media, MS starts to predominate beyond a few $l_s$. To cope with the fundamental issue of MS, several approaches have been proposed in order to enhance the SS contribution drowned into a predominant MS background. The first option is to spatially discriminate SS and MS as performed in confocal microscopy \cite{Pawley,Ntziachristos} or two-photon microscopy \cite{Theer}. The second option consists in separating SS from MS photons \coraa{by means of time gating \cite{Fujimoto2,Alfano,Alfano2}}. Probably, the most widely employed coherent time-gated technique is optical coherence tomography (OCT) \cite{fujimoto,fujimoto3,Fercher}, which is the analogous to ultrasound imaging. It combines scanning confocal microscopy with coherent heterodyne detection \cite{oct0}. OCT has drastically extended the imaging-depth limit compared to conventional microscopy. Nevertheless, its ability of imaging soft tissues remains typically restricted to a depth of 1 mm \cite{oct,Dunsby}. 

Inspired by previous works in ultrasound imaging through strongly scattering media \cite{aubry,aubry2,shahjahan}, we propose a matrix approach of optical imaging to push back the fundamental limit of MS. Experimentally, this approach, referred to as \textit{smart-OCT}, relies on the measurement of a time-gated reflection matrix from the scattering medium. Unlike previous works \cite{choi2,Kang}, the reflection matrix is here directly investigated in the focal plane on a point-to-point basis \cite{robert,robert2}. An input-output analysis of the reflection matrix allows to get rid of most of the MS contribution. Iterative time-reversal \cite{prada,prada2,popoff3} is then applied to overcome the residual MS contribution as well as the aberration effects induced by the turbid medium itself. \cora{A proof-of-concept experiment demonstrates imaging through a strongly scattering paper layer from which only one reflected photon over 1000 billion is associated to a SS event from the object hidden behind it. In a second experiment, our approach is successfully applied to optical imaging through a thick layer of biological tissues}. Compared to OCT and related methods \cite{Kang}, we show both theoretically and experimentally an extension of the imaging-depth limit by a factor two. This means a multiplication by a factor two of the sensitivity in dB compared to existing OCT systems.

\section*{RESULTS}

\subsection*{Measuring the time-gated reflection matrix}

\begin{figure*}[!ht]
\center
\includegraphics[width=17cm]{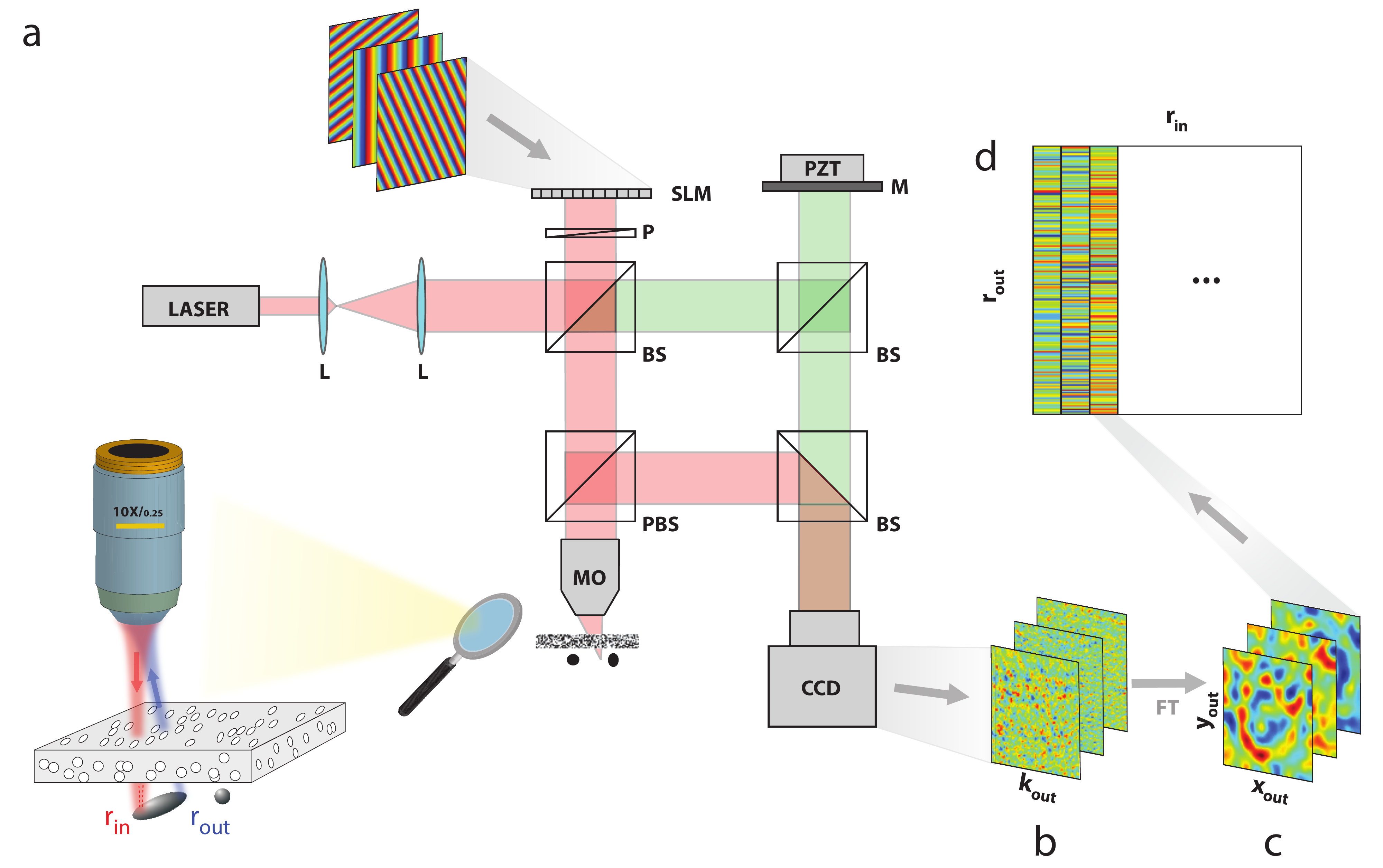}
\caption{\textbf{Measuring the time-gated reflection matrix}. (\textbf{A}), Experimental set up: P: polarizer, MO: microscope objective, BS: beam splitter, PBS: polarized beam splitter, SLM : spatial light modulator, PZT: piezo phase shifter, M: Mirror. A femtosecond laser beam (center wavelength: 810 nm, bandwidth: 40 nm) is shaped by an SLM acting as a diffraction grating. A set of incident plane waves is thus emitted from the SLM and focused at a different position in the focal plane of a MO (NA=0.25). The backscattered wave-field is collected through the same MO and interferes with a reference beam on a CCD camera. The latter one is conjugated with the back focal plane of the MO. The amplitude and phase of the wave-field is recorded by phase shifting interferometry \cite{popoff}. The time of flight $t$ is controlled by the length of the interferometric arm and is matched with the position of the focal plane. (\textbf{B}), For each input focusing point $\mathbf{r_{in}}$, a reflected wave-field $R(\mathbf{r_{in}},\mathbf{k_{out}})$ is recorded in the $\mathbf{k}$-space. (\textbf{C}), A 2D Fourier transform yields the wave-field in the real space, $R(\mathbf{r_{in}},\mathbf{r_{out}})$, where $\mathbf{r_{out}}$ represents the output focusing point in the focal plane. (\textbf{D}), For each incident focusing point $\mathbf{r_{in}}$, the recorded wave-field is stored along a column vector. The set of column vectors finally form the reflection matrix $\mathbf{R}=[R(\mathbf{r_{in}},\mathbf{r_{out}})]$.}
\label{fig1}
\end{figure*}

The smart-OCT approach is based on the measurement of a time-gated reflection matrix $\mathbf{R}$ from the scattering sample. Until now, optical transmission/reflection matrices have always been measured in the $\mathbf{k}$-space (plane wave basis) \cite{popoff,choi,choi2,Kang}. Here, inspired by previous studies in acoustics \cite{robert,robert2}, the reflection matrix is directly investigated in the real space (point-to-point basis). The experimental set up and procedure are described in Fig.~\ref{fig1}. A set of reflection coefficients $R(\mathbf{r_{out}},\mathbf{r_{in}})$ are measured between each point of the focal plane identified by the vectors $\mathbf{r_{in}}$ at the input and $\mathbf{r_{out}}$ at the output. These coefficients form the reflection matrix $\mathbf{R}$. In the experiments described below, the reflection matrix is measured over a field-of-view of $40 \times 40$ $\mu$m$^2$ mapped with $289$ input wave-fronts. \\

\subsection*{Reflection matrix in the single scattering regime}
\begin{figure*}[!ht]
\center
\includegraphics[width=17cm]{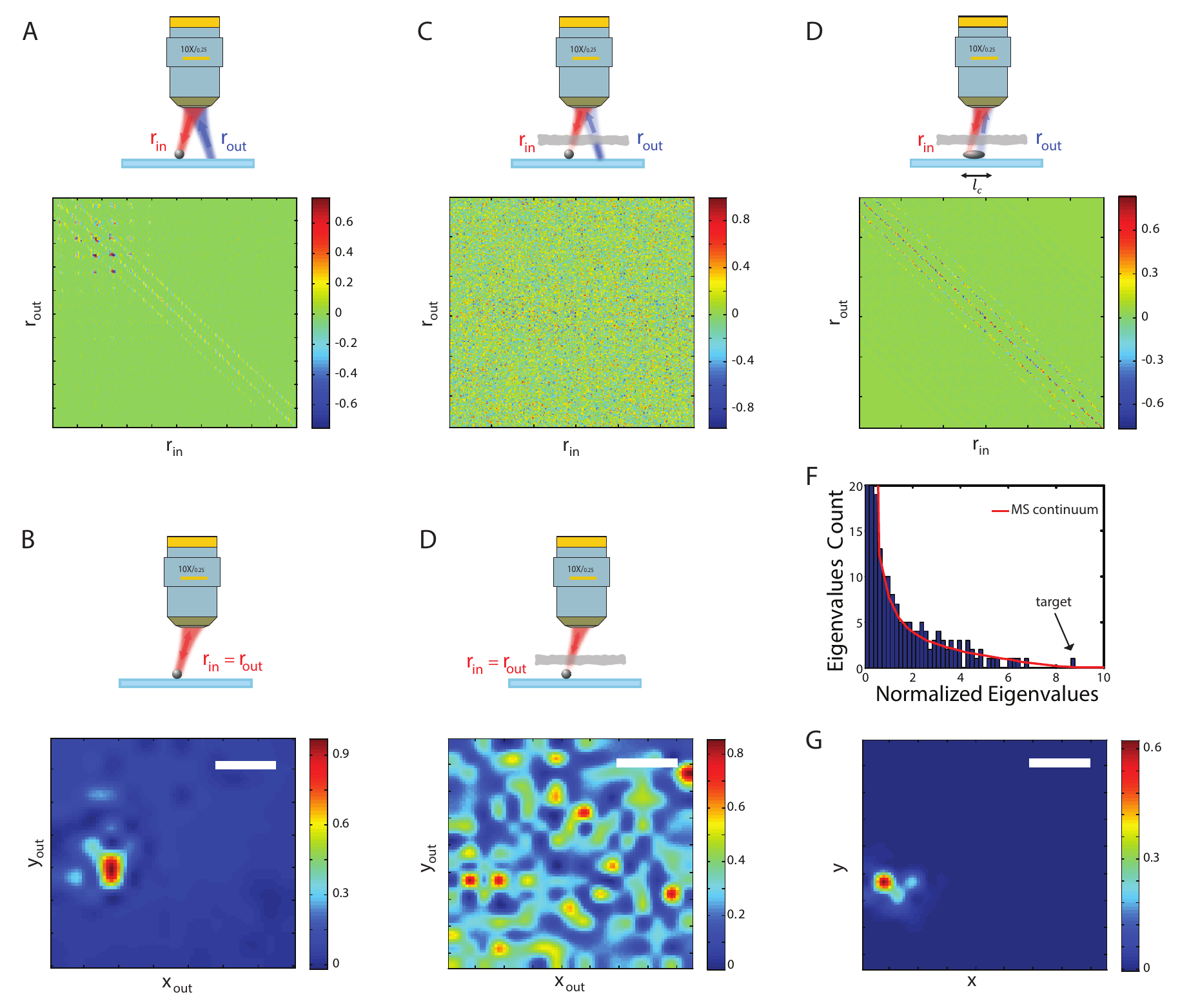}
\caption{ \textbf{Target detection in the deep multiple scattering regime}. (\textbf{A}) Reference time-gated reflection matrix $\mathbf{R_0}$ measured for a ZnO bead deposited on a glass slide. (\textbf{B}) \textit{En-face} OCT image deduced from $\mathbf{R_0}$ [equation~(\ref{eq1})]. (\textbf{C}) Time-gated reflection matrix $\mathbf{R}$ in presence of a strongly scattering layer ($L=12.25$ $l_s$). (\textbf{D}) \textit{En-face} OCT image deduced from $\mathbf{R}$ [equation~(\ref{eq1})]. (\textbf{E}) Single scattering matrix $\mathbf{R_S}$ deduced from $\mathbf{R}$ with $l_c=5$ $\mu$m [equation~(\ref{Rs})]. (\textbf{F}) Eigenvalue histogram of $\mathbf{R_SR_S^{\dag}}$ compared to the random matrix prediction (red line): The largest eigenvalue $\sigma_1^2$ clearly emerges from the MS noise. (\textbf{G}) Smart-OCT image deduced from the first eigenstate of $\mathbf{R_S}$. Scale bar: 10 $\mu$m.
}
\label{fig2}
\end{figure*}

Figure~\ref{fig2}A displays a reference reflection matrix $\mathbf{R_0}$ measured for a ZnO bead of diameter $d=10$ $\mu$m deposited on a microscope slide. $\mathbf{R_0}$ contains two main contributions: 
\begin{itemize}
\item The specular echo from the glass slide that emerges throughout the diagonal of $\mathbf{R_0}$\\
\item The strong bead echo that arises for positions $\mathbf{r_{in}} \sim \mathbf{r_{b}}$, with $\mathbf{r_{b}}$ the position of the bead in the focal plane. Because the bead diameter $d$ is larger than the width $\delta$ of the point spread function, the bead also emerges along off-diagonal elements for which $|\mathbf{r_{out}}-\mathbf{r_{in}}| \leq d$.
\end{itemize}
\cora{The single scattering contribution thus only emerges along the diagonal and closed-diagonal elements of $\mathbf{R_0}$. This is accounted for by the fact that a singly-scattered wave-field can only come from points illuminated by the incident focal spot.} A time-gated \textit{confocal} image can be deduced from $\mathbf{R_0}$ by only considering its diagonal elements, such that
\begin{equation} 
\label{eq1}
I_0(\mathbf{r})= {\left  |R_0 \left ( {\mathbf{r}} , {\mathbf{r}} \right) \right  |}^2
\end{equation}
The corresponding image displayed in Fig.~\ref{fig2}B is equivalent to a \textit{en-face} OCT image. Not surprisingly, it shows a clear image of the target on the microscope glass slide. 

\subsection*{Reflection matrix in the deep multiple scattering regime}

In a second experiment, a stack of two paper sheets is placed between the MO and the target bead [see Fig.~\ref{fig1}]. The thickness of each sheet is of 82 $\mu$m, hence an overall thickness of $L=164$ $\mu$m for the scattering layer. The distance between the front-surface of the scattering layer and the target is $F\simeq 1$ mm. The scattering and transport mean free paths in the paper sheet have been measured and estimated to be $l_s \sim 13.4$ $\mu$m and $l_t \sim 19.9$ $\mu$m, respectively [see Supplementary Section I]. This yields an optical thickness $L \sim 12.25 l_s \sim 8.2 l_t$. The ballistic wave has to go through 24.5 $l_s$ back and forth, thus undergoing an attenuation of $\exp(-24.5) \sim 2 \times 10^{-11}$ in intensity. The single-to-multiple scattering ratio (SMR) of the reflected wave-field is estimated to be $10^{-12}$ in the MO back-focal plane [see Supplementary Section II].  For an incident plane wave, it means that only 1 scattered photon over 1000 billion is associated to a SS event from the target. As shown in Supplementary Fig.~\ref{fig0bis}, the target is far to be detectable and imaged in this experimental configuration, whether it be by conventional microscopy (SMR $\sim 10^{-10}$), confocal microscopy (SMR $\sim 10^{-8}$) or by OCT (SMR $\sim 10^{-5}$). This experimental situation is thus particularly extreme, even almost desperate, for a successful imaging of the target.

Figure~\ref{fig2}C displays the reflection matrix $\mathbf{R}$ measured in presence of the scattering layer. Contrary to the SS contribution \cora{that emerges along diagonal and closed-diagonal elements of $\mathbf{R}$} [Fig.~\ref{fig2}A], MS randomizes the directions of light propagation and gives rise to a random reflection matrix \cite{aubry}. Nevertheless, one can try to image the target by considering the diagonal of $\mathbf{R}$ [equation~(\ref{eq1})]. The corresponding \textit{en-face} OCT image is shown in Fig.~\ref{fig2}D. As theoretically expected [see Supplementary Fig.~\ref{fig0bis}], MS still predominates despite confocal filtering and coherence time-gating. An image of speckle is thus obtained without any enhancement of the intensity at the expected target location [see the comparison with the reference image in Fig.~\ref{fig2}B]. \\

\subsection*{Matrix approach dedicated to target detection in the deep multiple scattering regime}

An alternative route is now proposed to image and detect the target behind the scattering layer. 
The smart-OCT approach first consists in filtering the multiple-scattering contribution in the measured reflection matrix $\mathbf{R}$. 
To that aim, the $\mathbf{R}$-matrix is projected on a characteristic SS matrix $\mathbf{S}$, whose elements are given by 
\begin{equation}
\label{s}
S(\mathbf{r_{out}},\mathbf{r_{in}})=\exp \left (-{|\mathbf{r_{in}}-\mathbf{r_{out}}|^2}/{l_c^2} \right )
\end{equation}
$l_c$ is a tunable parametric length that accounts for the fact that the ballistic signal does not only emerge along the diagonal of the $\mathbf{R}-$matrix but also along off-diagonal elements [Fig.~\ref{fig2}A]. $l_c$ is governed by two factors:
\begin{itemize}
\item The coherence length of the ballistic wave-field in the focal plane: In addition to ballistic attenuation and MS, the scattering layer also induces aberrations that degrades the focusing quality of the ballistic wave-front and enlarge the point spread function of the imaging system.
\item The size of the target: As shown by Fig.~\ref{fig2}A, the target signal does not only emerge along the diagonal elements of $\mathbf{R}$ in absence of the scattering layer. This is accounted for by the size of the target which is larger than the resolution cell.
\end{itemize}
Mathematically, the projection of $\mathbf{R}$ can be expressed as an Hadamard product with $\mathbf{S}$,
\begin{equation}
\label{Rs}
\mathbf{R_S}=\mathbf{R} \circ \mathbf{S},
\end{equation}
\cora{which, in term of matrix coefficients, can be written as
\begin{equation}
\label{Rs2}
R_S(\mathbf{r_{out}},\mathbf{r_{in}})=R(\mathbf{r_{out}},\mathbf{r_{in}}) \times S(\mathbf{r_{out}},\mathbf{r_{in}}).
\end{equation}}
\cora{This mathematical operation thus consists in keeping the diagonal and closed-diagonal coefficients of $\mathbf{R}$ where the SS contribution arises and filtering the off-diagonal elements of $\mathbf{R}$ mainly associated with the MS contribution. It can be seen as a digital confocal operation with a virtual pinhole mask of size $l_c$ \cite{psaltis}. In the present experiment,}  a SS matrix $\mathbf{R_S}$ is deduced from $\mathbf{R}$ by considering $l_c=5$ $\mu$m. The result is displayed in Fig.~\ref{fig2}E. $\mathbf{R_S}$ contains the SS contribution as wanted plus a residual MS contribution [see the comparison with the reference matrix in Fig.~\ref{fig2}A]. This term persists because MS signals also arise along and close to the diagonal of $\mathbf{R}$. \cora{Compared to a single/multiple scattering separation performed in the far-field \cite{aubry2,Kang}, a single scattering projection in a point-to-point basis [Eq.\ref{Rs}] is much more flexible since the tunable parameter $l_c$ can be adapted as a function of the aberration level or the expected target size.}

Once this SS matrix is obtained, one can apply the DORT method (French acronym for Decomposition of the Time Reversal Operator). Initially developed for ultrasound \cite{prada,prada2}, the DORT method takes advantage of the reflection matrix to focus iteratively by time reversal processing on each scatterer of a multi-target medium \cite{popoff3}. Mathematically, the time-reversal invariants can be deduced from the eigenvalue decomposition of the time reversal operator $\mathbf{R}\mathbf{R^{\dag}}$ or, equivalently, from the singular value decomposition of $\mathbf{R}$ (the superscript $\dag$ stands for transpose conjugate). A one-to-one association between each eigenstate of $\mathbf{R}$ and each scatterer does exist. On the one hand, the eigenvectors of $\mathbf{R}$ allow selective focusing and imaging of each scatterer. On the other hand, the associated eigenvalue directly yields the scatterer reflectivity. Nevertheless, this one-to-one association is only valid under a SS approximation. Hence the DORT method cannot be applied to the raw matrix $\mathbf{R}$ since it contains an extremely predominant MS contribution. The trick here is to take advantage of the SS matrix $\mathbf{R_S}$. 

A singular value decomposition (SVD) of $\mathbf{R_S}$ is performed. It consists in writing $\mathbf{R_S} = \mathbf{U\Sigma V^{\dag}}$. $\mathbf{\Sigma}$ is a diagonal matrix containing the real positive singular values $\sigma_{i}$ in a decreasing order $\sigma_1>\sigma_2> \cdots >\sigma_N$. The square of the singular values, $\sigma_i^2$, correspond to the eigenvalues of $\mathbf{R_S}\mathbf{R_S^{\dag}}$. $\mathbf{U}$ and $\mathbf{V}$ are unitary matrices whose columns correspond to the input and output singular vectors $\mathbf{U_i}$ and $\mathbf{V_i}$, respectively. Fig.~\ref{fig2}F displays the histogram of the eigenvalues $\sigma_{i}^2$ normalized by their average. It is compared to the distribution that would be obtained in a fully multiple scattering regime [see Supplementary Section III]. The histogram of $\sigma_i^2/<\sigma_i^2>$ in Fig.~\ref{fig2}F follows this distribution except for the largest eigenvalue $\sigma_1^2$. The latter one is actually beyond the superior bound of the MS continuum of eigenvalues. This means that the first eigenspace is associated to the target \cite{aubry2,shahjahan}. The combination of the first input and output singular vectors, $ | \mathbf{U_1} \circ \mathbf{V_1} |$, forms the smart-OCT image displayed in Fig.~\ref{fig2}G. The image of the target is nicely recovered. The comparison with the \textit{en-face} OCT image displayed in Fig.~\ref{fig2}D unambiguously demonstrates the benefit of smart-OCT in detecting a target in the deep MS regime ($L=12.25$ $l_s$). Note that the target image does not match exactly with the reference image [Fig.~\ref{fig2}B]. This difference can be accounted for by residual aberration effects induced by the scattering layer itself.

The theoretical study developed in the Supplementary Section II confirms this experimental result. In the conditions of the reported experiment, the imaging thickness-limit (SMR$\sim1$) is found to be of 1.5$l_s$ for conventional microscopy, 3.5$l_s$ for confocal microscopy and 7$l_s$ for OCT. This explains the failure of OCT in detecting the target [Fig.~\ref{fig2}D]. On the contrary, the predicted imaging-thickness limit is of 12$l_s$ for the smart-OCT approach. This accounts for the successful detection of the target in our experimental configuration [Fig.~\ref{fig2}G].

\subsection*{Imaging in the deep multiple scattering regime}
\begin{figure*}[!ht]
\center
\includegraphics[width=17cm]{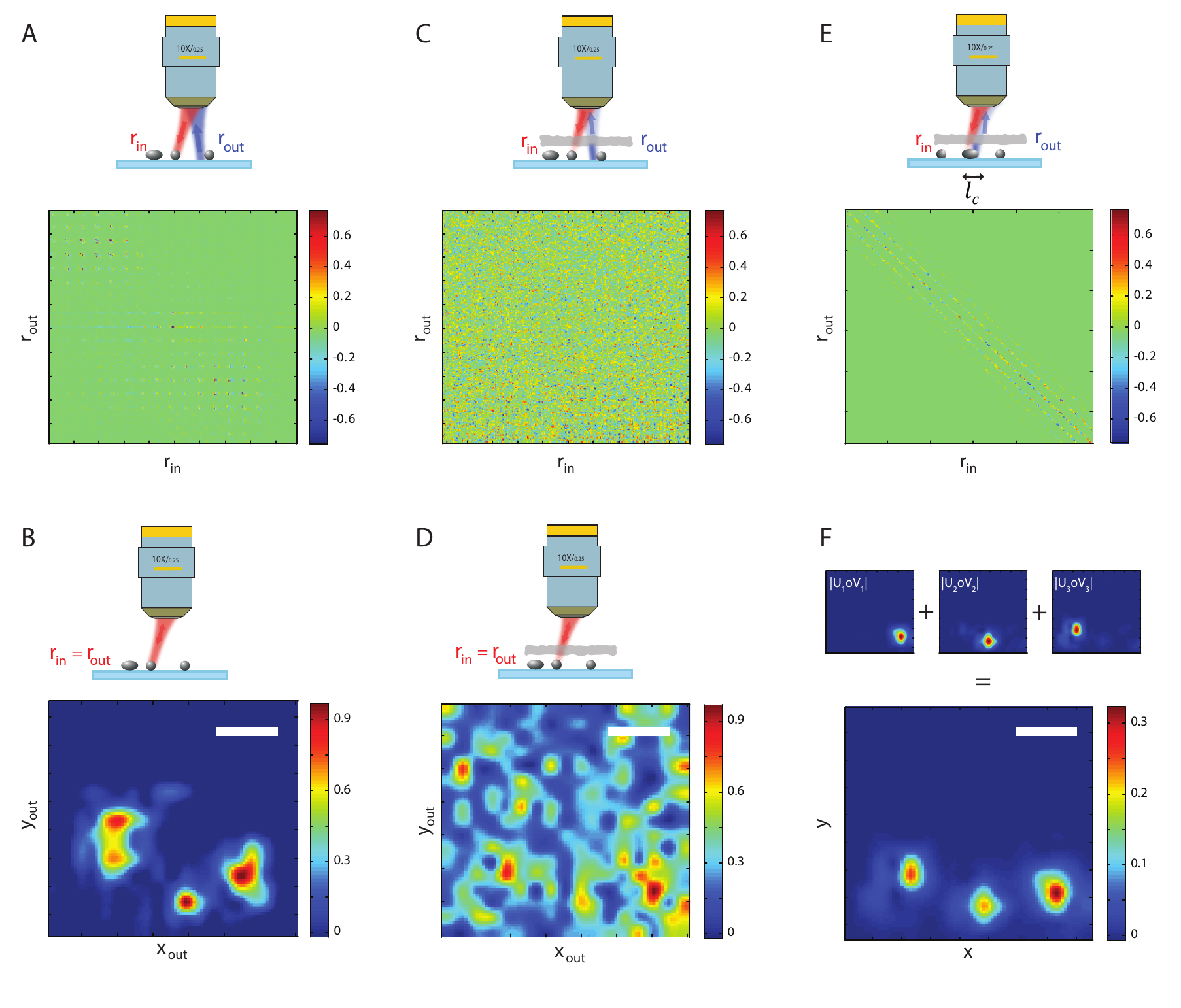}
\caption{\textbf{Imaging in the deep multiple scattering regime}. (\textbf{A}) Reflection matrix $\mathbf{R}$ associated to three 5-$\mu$m diameter ZnO beads deposited on a glass slide. (\textbf{B}), Time-gated confocal image of the three beads in absence of the scattering layer. (\textbf{C}) Reflection matrix $\mathbf{R}$ measured in presence of a scattering layer ($L=6.2$ $l_s$) placed before the three beads. (\textbf{D}) Time-gated confocal image in presence of the scattering layer. (\textbf{E}) Single scattering matrix $\mathbf{R_s}$ built from $\mathbf{R}$ using Eq.\ref{Rs}. (\textbf{F}) The three first eigenstates of $\mathbf{R_s}$, $|\mathbf{U_i}\circ \mathbf{V_i}|$, are combined to yield the smart-OCT image of the three beads in presence of the scattering layer. Scale bar: 10 $\mu$m.}
\label{fig3}
\end{figure*}
From the previous experiment, one could say that the smart-OCT approach is only a single target detection method dedicated to the deep MS regime. To demonstrate that we can go beyond the detection and imaging of a single target, a configuration with three 5$-\mu$m-diameter ZnO beads deposited on a microscope slide is investigated. Figure~\ref{fig3} (A and B) display the time-gated reflection matrix $\mathbf{R_0}$ and the corresponding \textit{en-face} OCT image in absence of any scattering layer. Both figures will serve as reference in the following. A paper sheet is then placed between the MO and the focal plane. The corresponding reflection matrix $\mathbf{R}$ is shown in Fig.~\ref{fig3}C. Not surprisingly, it displays a random feature characteristic of a predominant MS contribution. The \textit{en-face} OCT image built from the diagonal elements of $\mathbf{R}$ is displayed in Fig.~\ref{fig3}D. Again the MS speckle prevents from a clear and unambiguous detection of the three targets [see the comparison with Fig.~\ref{fig3}B]. The MS filter is applied to the raw matrix $\mathbf{R}$ [equation~(\ref{Rs}), $l_c=5$ $\mu$m], yielding a SS matrix $\mathbf{R_S}$ displayed in Fig.~\ref{fig3}E. Iterative time-reversal is then performed. The three first eigenstates of $\mathbf{R_S}$ are displayed in Fig.~\ref{fig3}F. A comparison with the reference image in Fig.~\ref{fig3}B highlights the one-to-one association between each bead and each of these eigenstates.  The combination of these eigenstates weighted by the corresponding eigenvalue finally provide the smart-OCT image. The comparison with the \textit{en-face} OCT image [Fig.~\ref{fig3}D] demonstrates the success of the smart-OCT approach for imaging in the deep MS regime.

\subsection*{Imaging through thick biological tissues}

\begin{figure*}[!ht]
\center
\includegraphics[width=17cm]{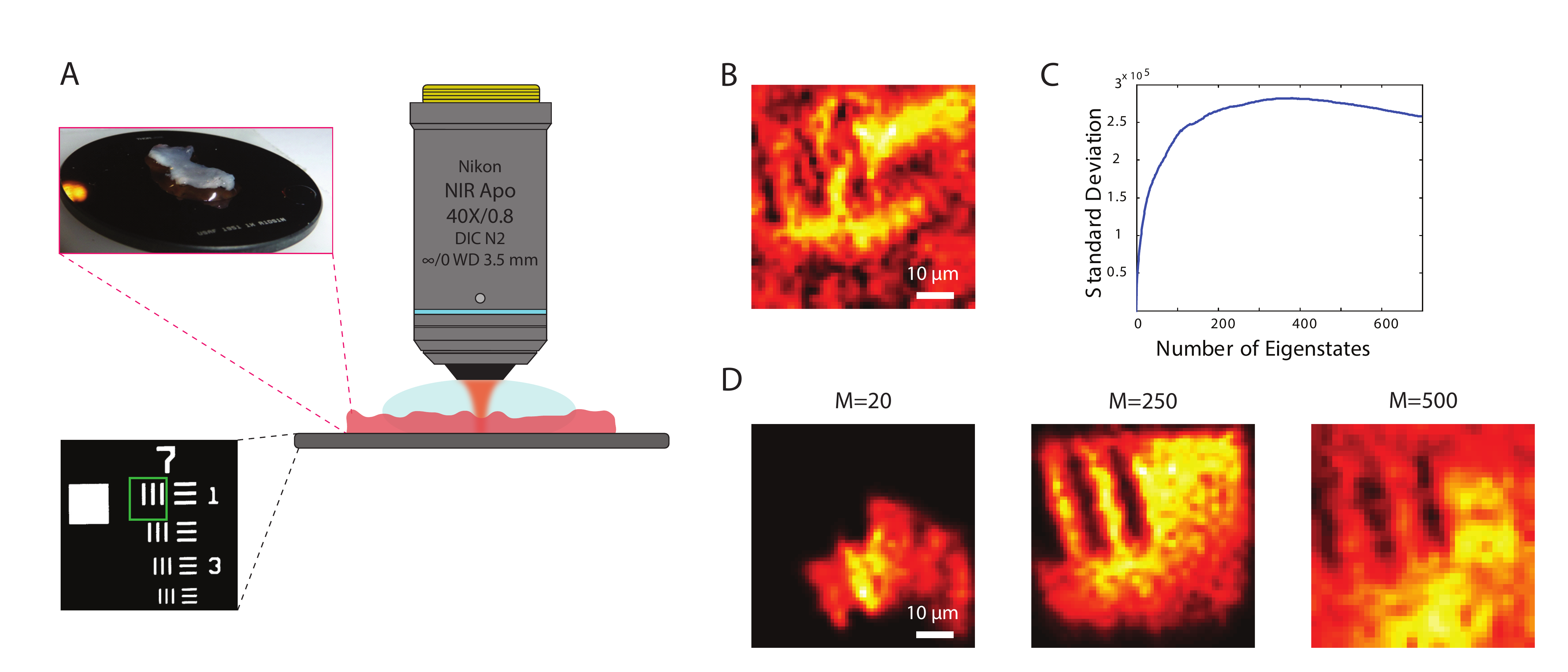}
\caption{\textbf{Imaging through thick biological tissues}. (A) Schematic of the experimental configuration. A immersion microscope objective image a positive resolution target USAF 1951 with a 800 $\mu$m thick layer of rat intestine on top of it. The region of interest is surrounded by a green square in the bottom inset. (B) \textit{En-face} OCT image of the resolution target. (C) Standard deviation of the smart-OCT image as a function of the number $M$ of eigenstates of $\mathbf{R_S}$ considered. (D,E,F) Smart-OCT images of the resolution target obtained from the 20, 250 and 500 first eigenstates of $\mathbf{R_S}$, respectively. }
\label{fig4}
\end{figure*}
Following this experimental proof-of-concept, we now apply our approach to the imaging of an extended object through biological tissues. A positive USAF 1951 resolution target placed behind a 800 $\mu$m-thick layer of rat intestine tissues is imaged through an immersion objective (Olympus, $\times$40, NA=0.8) [see Fig.~\ref{fig4}A]. The reflection matrix $\mathbf{R}$ is measured over a field-of-view of $60 \times 60$ $\mu$m$^2$ [see the green square in Fig.~\ref{fig4}A] with $961$ input wave-fronts. The diagonal of $\mathbf{R}$ yields the \textit{en-face} OCT image displayed in Fig.~\ref{fig4}B. Due to the aberration effects and multiple scattering events induced by the biological tissues, the three bars of the USAF target cannot be recovered. A comparable result is obtained if the DORT method is directly applied to the raw matrix $\mathbf{R}$ [See Supplementary Fig.~\ref{FigS3}]. To overcome these detrimental effects, the MS filter is applied to the raw matrix $\mathbf{R}$ [equation~(\ref{Rs}), $l_c=8$ $\mu$m], yielding a SS matrix $\mathbf{R_S}$. Iterative time-reversal is then performed. In previous experiments, the object to image consisted in one or a few beads. This sparsity implied that only few eigenstates were needed to recover the image of the beads. In the present case, the USAF target is an extended object. It is thus associated with a large number $M$ of eigenstates, $M$ scaling as the number of resolution cells contained in the object \cite{robert3}. To estimate the rank $M$ of the object, one can compute the standard deviation of the image, $\left | \sum_n \sigma_n \mathbf{U_n} \circ \mathbf{V_n} \right | $, as a function of the number $n$ of eigenstates considered for the imaging process. The result is displayed in Fig.~\ref{fig4}C. A maximum standard deviation is found for $M \sim 250$ eigenstates. The corresponding image is displayed in Fig.~\ref{fig4}E. The three bars of the USAF target are nicely recovered and the comparison with the \textit{en-face} OCT image [Fig.~\ref{fig4}B] is striking. This experimental result demonstrates the benefit of our approach for deep-tissue imaging. To illustrate the importance of a correct determination of $M$, we also show for comparison the images built from the 20 and 500 first eigenstates of $\mathbf{R}$ [see Fig.~\ref{fig4}(D and F), respectively]. On the one hand, considering too few eigenstates only provides a partial imaging of the field of view [Fig.~\ref{fig4}D]. On the other hand, considering too many eigenstates blurs the image since the weakest eigenvalues are mainly associated with the multiple scattering background [Fig.~\ref{fig4}F]. 

\section*{DISCUSSION}
\begin{figure}[!ht]
\center
\includegraphics[width=8.5cm]{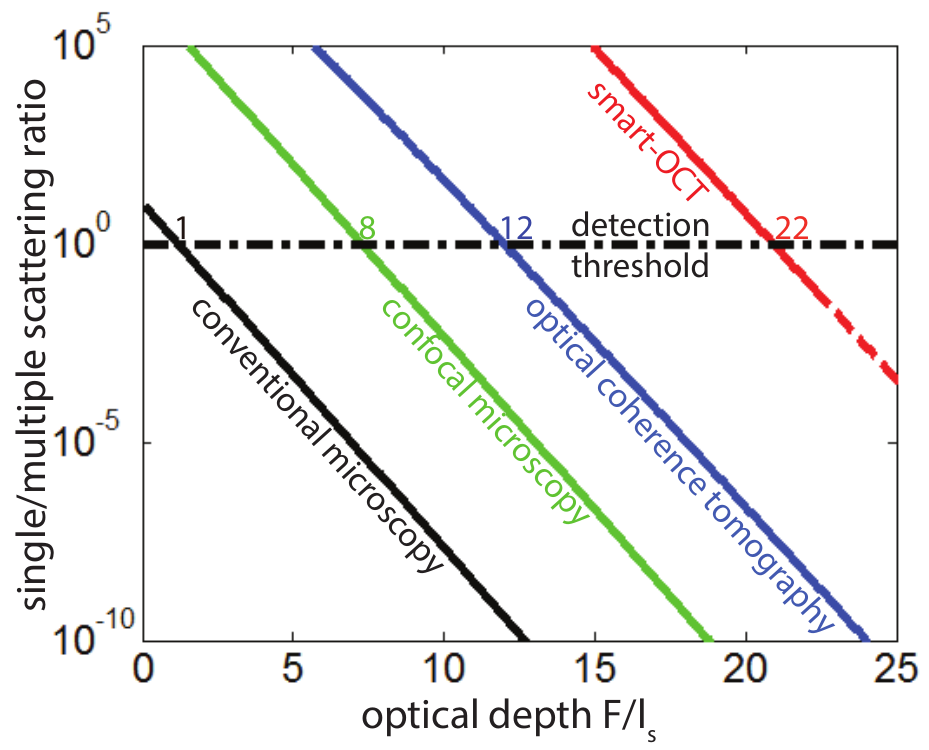}
\caption{\textbf{Imaging-depth limit in human soft tissues}. This graph compares the single-to-multiple scattering ratio (SMR) expected for conventional microscopy (black line), confocal microscopy (green line), OCT (blue line) and smart OCT (red line) as a function of the optical depth $F/l_s$. The y-axis is in log-scale. These curves have been computed from the theoretical study developed in the Supplementary Section II considering experimental parameters typical of full-field OCT \cite{Assayag}. The detection threshold (SMR$\sim1$, black dashed horizontal line) yields an imaging depth limit of $\sim 1l_s$ for conventional microscopy, $\sim8l_s$ for confocal microscopy, $\sim 12 l_s$ for OCT and $\sim 22 l_s$ for smart-OCT.}
\label{fig5}
\end{figure}
These experimental results demonstrate the benefit of the smart-OCT approach for optical imaging in strongly scattering media. In the Supplementary Section II, this superiority is confirmed by the theoretical investigation of the imaging-depth limit derived for different imaging techniques (conventional/confocal microscopy, OCT, smart-OCT). In view of applications to deep tissue imaging, Fig.~\ref{fig5} shows the SMR evolution versus the optical depth $F$ in biological tissues. The experimental parameters chosen for this figure are those typically encountered in full-field OCT \cite{Assayag} and are provided in Supplementary Tab.S1. The detection threshold is set at a SMR of 1. The imaging-depth limit expected in tissues is of 1$l_s$ for conventional microscopy, 8$l_s$ for confocal microscopy and 12$l_s$ for OCT. \cora{The latter value is in agreement with the imaging-depth limit recently reached by Kang et al. \cite{Kang} with an optical technique similar to OCT. It actually combines coherence time gating with a spatial input-output correlation of waves from the far-field that allows a confocal discrimination of reflected photons. On the contrary, our approach goes beyond OCT as it also involves a subsequent iterative time-reversal processing of the reflection matrix. It results in an additional gain in SMR that scales with $N$, the number of input wave-fronts (see Supplementary Section II). This leads to an imaging-depth limit of 22$l_s$ in Fig.~\ref{fig5}, hence multiplying by almost two the current OCT limit. Such an imaging improvement is drastic if we keep in mind that the ballistic contribution decreases by a factor $\exp(-2L/l_s)$ in a reflection configuration. The smart-OCT approach is thus particularly suited for ultra-deep tissue imaging. Of course, a trade-off will have to be made in practice between the imaging depth and the measurement time that also scales linearly with $N$. 
Note also that the imaging depth can be limited by the dynamic range of the CCD camera. For instance, a dynamic range of $75$ dB would be required to reach the theoretical imaging depth-limit in the experimental conditions of Fig.~\ref{fig5}  (see Supplementary Tab.~\ref{table}).}

\cora{A second point we would like to discuss is the resolution and sectioning capabilities of our approach. On the one hand, as the image is built from singly-scattered photons, the transverse resolution is only diffraction-limited and does not depend on the penetration depth. On the other hand, the coherent time-gated detection scheme provides an axial resolution $\delta z$ that is only governed by the coherence time $\tau_c$ of the light source: $\delta z \sim c \tau_c/(2n)$, with $n$ the refractive index of the medium. In the present experiment, the coherence time of the femtosecond laser is of 50 fs. It yields an axial resolution of 5 $\mu$m in biological tissues ($n\sim 1.4$). By measuring a set of reflection matrices at successive depths, a 3D image of the sample can thus be obtained as in conventional OCT with the great advantage that the penetration depth is multiplied by a factor 2. To reach a better axial resolution, the measurement of the reflection matrix can be made under a simple white light illumination. A recent study has actually demonstrated the passive measurement of the time-dependent point-to-point reflection matrix from an incoherent illumination \cite{badon}. As already demonstrated in full-field OCT \cite{Assayag}, the temporal incoherence of the white-light source would provide an excellent axial resolution ($\delta z\sim$1 $\mu$m). Moreover, such a device would comply with the non-invasive, low-cost, speed and low-complexity specifications required for medical applications.}

In summary, this study proposes a matrix approach of light propagation dedicated to optical detection and imaging through complex media. The so-called smart-OCT approach combines a matrix discrimination of ballistic waves with iterative time-reversal. \cora{A first proof-of-concept experiment demonstrates the imaging of several micro-beads in the deep multiple scattering regime, whereas existing imaging techniques such as OCT are shown to fail. A second experiment demonstrates the diffraction-limited imaging of an extended object (USAF target) through a thick layer of biological tissues}. A theoretical investigation also demonstrates the significant superiority of our approach compared to confocal microscopy or OCT. In particular, an imaging-depth limit of 22 $l_s$ is predicted for smart-OCT in biological tissues, hence pushing back drastically the fundamental multiple scattering limit in optical imaging. 

\section*{MATERIALS AND METHODS}

The following components were used in the experimental set-up [Fig.~\ref{fig1}]: A femtosecond laser (FEMTOSECOND\textsuperscript{TM} FUSION\textsuperscript{TM} 20-400), a spatial light modulator (PLUTO NIR2, Holoeye), an objective lens (Olympus, $\times$10, NA=0.25), a CCD camera (DALSA Pantera 1M60) \cora{with a dynamic range of 60 dB. The incident light power is of 10 mW in the experiment. The radiant flux is thus of $10^6$ W.cm$^{-2}$ at the focal spot in free space. For each wave-front input, the complex reflected wave-field is extracted from four intensity measurements. Hence, the measurement of each line of the reflection matrix can be done at 15 Hz. According to the single-to-multiple scattering ratio, the reflected wave-field has to be averaged over a given number $n$ of measurements. In the imaging experiments through the paper layer [Figs.~\ref{fig2} and \ref{fig3}] and biological tissues [Fig.~\ref{fig4}], this number $n$ has been fixed to 32 and 5, respectively. Hence, the duration time for the recording of the reflection matrix has been of $\sim$10 min for the paper (289 input wave-fronts) and $\sim$5 min for biological tissues (961 input wave-fronts). The numerical post-processing of the reflection matrix (single scattering projection and iterative time-reversal) to get the final image only takes a few seconds.}


%

\section*{ACKNOWLEDGMENTS}
\noindent \textbf{Fundings} The authors are grateful for funding provided by LABEX WIFI (Laboratory of Excellence within the French Program Investments for the Future, ANR-10-LABX-24 and ANR-10-IDEX-0001-02 PSL*). A. B. acknowledges financial support from the French ``Direction Générale de l'Armement''(DGA). D. L. acknowledges financial support from LABEX WIFI and European Research Council (ERC Synergy HELMHOLTZ). G. L. and A. A. would like to acknowledge funding from High Council for Scientific and Technological Cooperation between France and Israel under reference P2R Israel N$^o$ 29704SC.\\
\noindent \textbf{Author Contributions} A.A. initiated and supervised the project. A.A. conceived the experiment. A.B. and D.L. built the experimental set up and performed the experiments. A.B., D.L. and A.A. analyzed the experiments. A.A. performed the theoretical study. A.B. and A.A. prepared the manuscript. All authors discussed the results and contributed to finalizing the manuscript.\\


\renewcommand{\thetable}{S\arabic{table}}
\renewcommand{\thefigure}{S\arabic{figure}}
\renewcommand{\theequation}{S\arabic{equation}}

\setcounter{equation}{0}
\setcounter{figure}{0}
\setcounter{page}{1}
\section*{SUPPLEMENTARY MATERIALS}

\subsection{Optical characterization of the paper layers}
White paper sheets are used as turbid media in our experiment. White paper is a strongly scattering medium with negligible absorption \cite{Hubbe}. In this section, we report on the measurement of the transport parameters that govern the diffusion of light across each paper sheet.

\subsubsection*{Scattering mean free path $l_s$}

\begin{figure}[h!]
\center
\includegraphics[width=8.5cm]{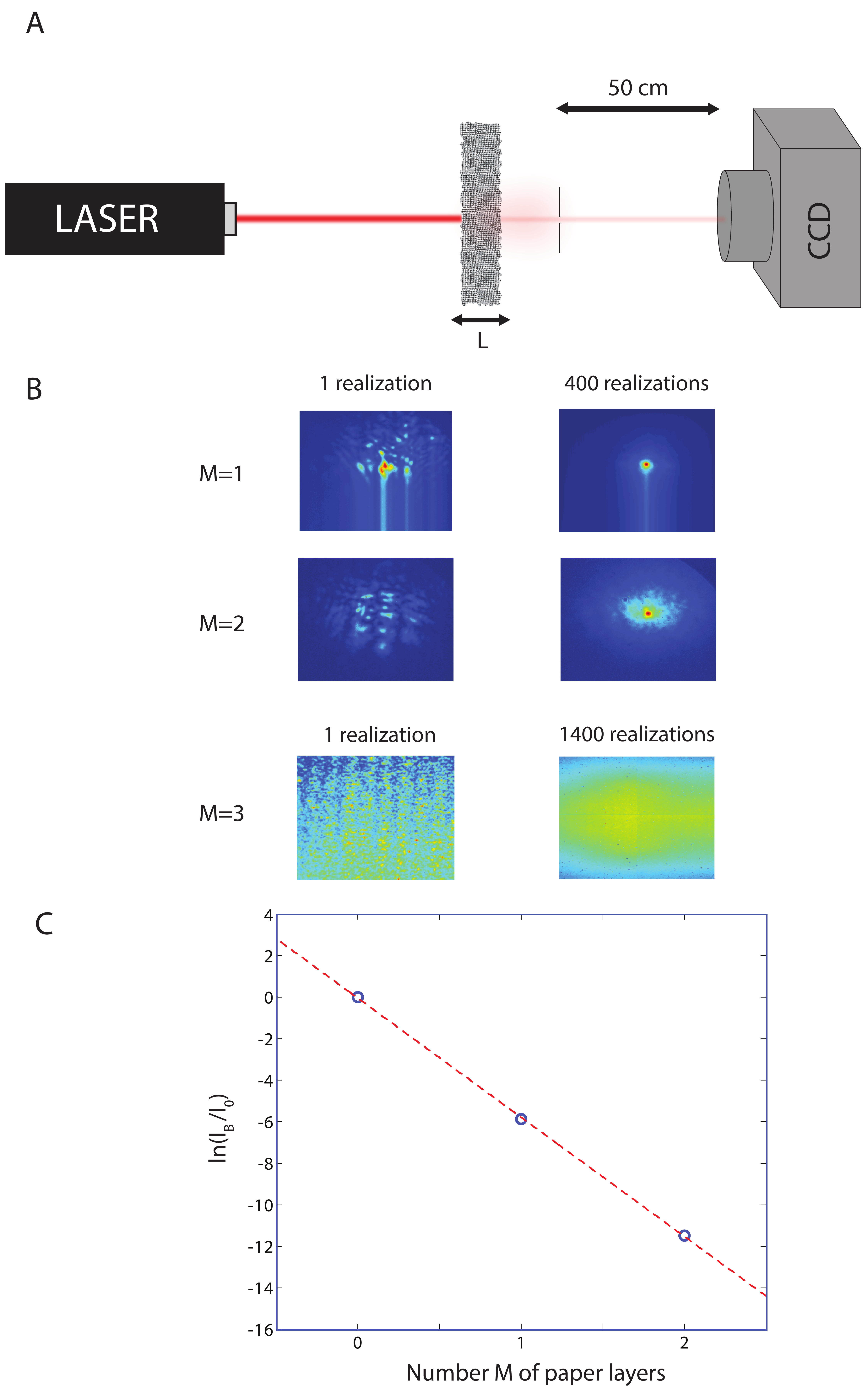}
\caption{\label{S1} \textbf{Measurement of the scattering mean free path in the paper. } (\textbf{A}) Experimental set up used for the measurement of the ballistic intensity $I_B$. (\textbf{B}) Left: Transmitted intensity recorded by the CCD camera for one realization of disorder. Right: Mean intensity obtained by averaging over 400/1400 realizations of disorder. The case of 1, 2 and 3 layers are displayed from top to bottom. (\textbf{C}) $\ln (I_B/I_0)$ as a function of the number of layers. A linear fit (red dashed line) of experimental points (blue circles) leads to an estimation of $l_s \simeq$13.4 $\mu$m (Eq.\ref{bal}).  }
\end{figure}
The ballistic component $I_B$ of the total transmitted intensity decays exponentially across a scattering layer of thickness $L$ \cite{akkermans},
\begin{equation}
\label{bal}
I_B=I_0\exp \left (-\frac{L}{l_s} \right )
\end{equation}
with $I_0$ the incident intensity. The scattering mean free path $l_s$ can be measured by investigating the attenuation of the ballistic light across an increasing number $n$ of paper sheets. To that aim, the experiment described in Fig.~\ref{S1}A has been performed. A collimated laser beam ($\lambda$=810 nm) illuminates the paper sheets. The transmitted intensity is the sum of the ballistic and the diffuse light. A pinhole is placed at the output of the medium to spatially discriminate the ballistic photons. The transmitted intensity is recorded on a CCD camera placed 50 cm behind the pinhole.

The transmitted intensity pattern is shown in Fig.~\ref{S1}B for configurations with 1, 2 and 3 sheets of paper. The left column represents a typical intensity pattern recorded for one realization of disorder. The right column represents the average of the transmitted intensity over a given number of disorder realizations. This is done by scanning spatially the surface of the scattering medium. For 1 and 2 layers, the ballistic component emerges on top of the diffusive halo. Its intensity can be estimated by subtracting the maximum intensity with the background intensity in the vicinity of the ballistic peak. For 3 layers of paper, the ballistic component cannot be revealed despite averaging over $1400$ realizations of disorder. This confirms that white paper is strongly scattering. The thickness $L'$ of each paper sheet has been measured with a precision caliper: $L' = 82$ $\mu$m. The optical thickness $L'/l_s$ of one paper sheet can be obtained by fitting linearly $ln(I_B/I_0)$ as a function of the number $M$ of layers [Fig.~\ref{S1}C]. We obtain $L'/l_s \simeq 6.1$. This leads to the following estimation of the scattering mean free path: $l_s\simeq$ 13.4 $\mu$m.

\subsubsection*{Transport Mean Free Path $l_t$}
\begin{figure}[h!]
\center
\includegraphics[width=8.5cm]{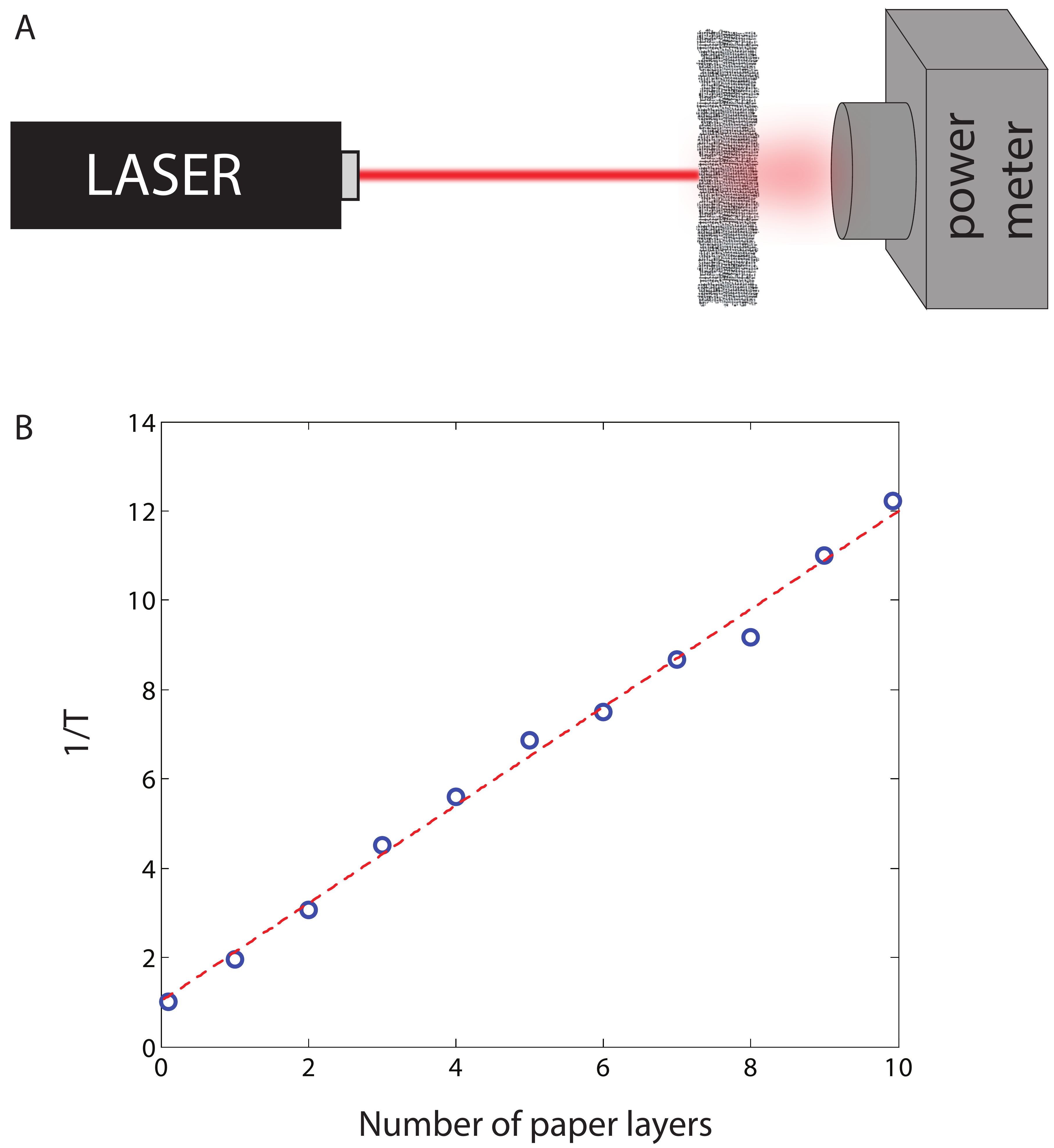}
\caption{\label{S2}\textbf{Measurement of the transport mean free path in the paper.} (\textbf{A}) Experimental set up used for the measurement of the transmission coefficient $T$. (\textbf{B}) Inverse of $T$ as a function the number $M$ of paper layers. A linear fit (red dashed line) of experimental points (blue circles) leads to an estimation of $l_t\simeq$19.9$\mu$m (Eq.\ref{ohm}).}
\end{figure}

Since $l_s<<L'$, most of the transmitted light is scattered several times while propagating through one sheet. For several sheets, the diffusion approximation becomes valid. The total transmission coefficient can then be written as follows \cite{durian},
\begin{equation}
\label{ohm}
T=\frac{z_0+l_t}{L+2z_0}
\end{equation}
with $z_0$ the distance from the scattering layer boundary at which the energy flux should cancel according to diffusion theory. In presence of a refractive index mismatch at the scattering madium boundaries, it can be expressed as \cite{Xhu}
\begin{equation}
z_0=\frac{2}{3}l_t\frac{1+R}{1-R}
\end{equation}
with $R$ the internal reflection coefficient. Considering a mean refractive index of 1.5 for the paper \cite{Hubbe},  $R=0.57$ \cite{Xhu} and $z_0 \simeq 2.4 l_t$.

The measurement of $l_t$ has been performed with the experiment described in Fig.~\ref{S2}A. A collimated laser beam illuminates the scattering sample and most of the transmitted light is collected with a powermeter. The transmission coefficient $T$ is measured for an increasing number $M$ of paper layers. Figure~\ref{S2}B displays $1/T$ as a function of M. A linear fit of experimental points leads to an estimation of $l_t$ using Eq.\ref{ohm}. With $z_0=2.4 l_t$, we find $l_t\simeq$19.9$\mu$m.

\subsection{Single-to-multiple scattering ratio for various imaging techniques} 

To address the issue of MS in optical imaging, it is important to establish theoretically the limit of existing imaging techniques in inhomogeneous media. To that aim, the relevant parameter is the single-to-multiple scattering ratio (SMR). In the following, this quantity is derived theoretically for several imaging techniques (conventional/confocal microscopy, OCT, smart-OCT) to highlight the gain in SMR provided by each of them. To that aim, we will consider the detection and imaging of an object embedded in or hidden behind a strongly scattering medium at a depth $F$. 

Let us first consider a conventional microscopy configuration. The target is placed in the focal plane of a microscope objective (MO). Under a plane wave illumination, the SMR of the backscattered wave-field in the back-focal plane of a MO is given by,
\begin{equation}
\label{SMR}
\mbox{SMR}=\frac{1}{4 \pi}\frac{ d\sigma/d{\Omega}}{W^2} \frac{ \exp({ -{2L}/{l_s}})}{\alpha} 
\end{equation}
with $d\sigma/d{\Omega}$, the differential scattering cross-section of the target, $W$, the field-of-view (FOV) of the optical system and $\alpha$, the static albedo of the scattering layer \cite{akkermans,Akkermans2}. Not surprisingly, SS is favored by the brightness of the target. However, the most important parameter here is the attenuation undergone by the ballistic wave across the scattering layer. Only a very tiny fraction of the incident energy, $\exp(-2L/l_s)$, is converted into the SS contribution used for imaging. In addition to the severe attenuation of the ballistic wave, the turbid medium gives rise to a speckle wave-field whose intensity is given by the static albedo $\alpha$ . For a scattering layer of thickness $L>>l_s$, its expression is given by \cite{akkermans}
\begin{equation}
\label{static_albedo}
\alpha \sim \frac{3}{8\pi} \left ( 1-\frac{2l_s}{L}\right ) 
\end{equation}

When one tries to image the target with conventional microscopy, the SMR in the conjugate image plane, referred to as SMR$_{m}$, can be expressed as follows
\begin{eqnarray}
\label{SMRm}
\mbox{SMR}_m &= & S \left ( W/\delta \right )^2 \mbox{SMR} \\
&= & \frac{S}{4 \pi}  \frac{ d\sigma/d{\Omega}}{\delta^2} \frac{ \exp({ -{2L}/{l_s}})}{\alpha} 
\end{eqnarray}
with $S$, the Strehl ratio ranging from 0 to 1 \cite{Mahajan}, $\delta=\lambda/(2NA)$, the resolution length of the imaging system and $NA$, the MO numerical aperture. Compared to conventional microscopy, the SMR is increased by a factor $N=(W/\delta)^2$ which corresponds the number of resolution cell in the FOV. Whereas the MS background results from the incoherent superimposition of $N$ independent speckle grains, the target image results from the constructive interference of the ballistic photons over the numerical aperture. However, the aberration undergone by the target ballistic wave-front across the scattering layer degrades the target focal spot and lowers its intensity. This is accounted for by the Strehl ratio $S$ in Eq.\ref{SMRm}. $S$ is directly proportional to the focusing parameter introduced by Mallard and Fink in the ultrasound imaging context \cite{mallart}. 

To cope with the fundamental issue of MS, several approaches have been proposed in order to enhance the SS contribution drowned into a predominant MS background. The first option is to spatially discriminate SS and MS as performed in confocal microscopy. Ideally the incoming radiation is focused to a single voxel and only light backscattering from that voxel is collected, allowing to reject a large number of multiply-scattered photons . However, scattered light can blur the focused beam outside the target volume and unwanted photons from other voxels can be scattered back into trajectories that will be collected by the microscope. Theoretically, the SMR provided by confocal microscopy, referred to as SMR$_c$, can be expressed as,
\begin{eqnarray}
\label{SMRc}
\mbox{SMR}_c &= & S \left ( W/\delta \right )^2 \mbox{SMR}_m =  S^2 \left ( W/\delta \right )^4 \mbox{SMR} \\
&= & \frac{S^2}{4 \pi)}  \left ( \frac{W}{\delta} \right )^2 \frac{ d\sigma/d{\Omega}}{\delta^2} \frac{ \exp({ -{2L}/{l_s}})}{\alpha} 
\label{SMRc2}
\end{eqnarray}
Compared to conventional microscopy, the SMR is increased a new time by the factor $ N$ which results from the coherent summation of the incident ballistic wave-front at the target location. The Strehl ratio $S$ in Eq.\ref{SMRc} accounts for the aberration effect undergone by this incident wave-front. 

The second way to enhance SS relatively to MS consists in discriminating SS from MS photons with coherence time gating. It allows to select the ballistic photons over a time window centered around their time of flight. Probably, the most widely employed coherence time-gated technique is optical coherence tomography (OCT), which combines scanning confocal microscopy with coherent heterodyne detection. The MS intensity is now given by a time-dependent albedo $\alpha(t)$ \cite{akkermans,Akkermans2}, with $t=2F/c$, the ballistic time. The SMR in OCT can be deduced from Eq.\ref{SMRc2} by substituting the static albedo $\alpha$ with the ratio $\Delta \omega/\alpha(t)$, with $\Delta \omega$ the bandwidth of the light source (inversely proportional to its coherence time $\tau_c$). The SMR in OCT, referred to as SMR$_t$, is thus given by,
\begin{eqnarray}
\label{SMRt}
\mbox{SMR}_{t} &= & \frac{S^2}{4 \pi}  \left ( \frac{W}{\delta} \right )^2 \frac{ d\sigma/d{\Omega}}{\delta^2} \frac{ \Delta \omega}{\alpha(t)} \exp({ -{2L}/{l_s}})\\
&= & \left [ \Delta \omega  {\alpha}/{\alpha(t)} \right ]   \mbox{SMR}_{c} 
\label{SMRt2}
\end{eqnarray}
Compared to confocal microscopy, the SMR in OCT is increased by a factor $\Delta \omega\alpha/\alpha(t)$ which accounts for the number of multiply scattered photons rejected by coherence time gating. The time-dependent albedo $\alpha(t)$ differs according to the imaging configuration. If the target is placed behind a scattering layer of thickness $L$, the following expression of $\alpha(t)$ should be considered \cite{Lagendijk} 
\begin{equation}
\label{albedo_t}
\alpha(t)\simeq \frac{\pi}{2}\frac{c(z_0+l_s)^2}{(L+2z_0)^3}\exp \left [ - \frac{\pi^2 Dt }{(L+2z_0)^2}\right ]
\end{equation}
where $D=cl_t/3$ is the diffusion constant that governs wave transport across the scattering layer. If the target is embedded within the scattering medium, the time-dependent albedo for a semi-infinite medium should be considered \cite{akkermans,Akkermans2}
\begin{equation}
\label{albedo_t2}
\alpha(t)\simeq \frac{c(z_0+l_s)^2}{(4\pi Dt)^{3/2}}
\end{equation}
According to the experimental configuration, the time-dependent albedo either displays a power law or an exponential decrease with the time-of-flight. In both cases, coherence time gating allows to drastically reject multiply-scaterred photons. 

In smart-OCT, in addition to confocal and coherence time gating operations, an eigenvalue decomposition of the single scattering reflection matrix $\mathbf{R_S}$ allows to get rid of the residual multiple scattering contribution. Whereas each ballistic echo emerges along one single eigenstate of $\mathbf{R_S}$ (the signal subspace), the incoherent MS wave-field emerges with the same probability along its $N$ eigenstates \cite{aubry2}. This implies an enhancement of the SMR by a factor $N=(W/\delta)^2$ compared to OCT [Eq.\ref{SMRt}]. Moreover, smart-OCT does not suffer from aberration issues in terms of detection. Time-reversal processing directly yields the distorted wave-front that compensates for the aberration effects induced by the scattering layer \cite{prada2,popoff3,aubry2}. The SMR ratio is thus independent of the Strehl ratio $S$. As a consequence, the SMR associated with smart-OCT, referred to as SMR$_s$, can be expressed as,
\begin{eqnarray}
\mbox{SMR}_s &= & S^{-2} \left ( W/\delta \right )^2 \mbox{SMR}_t \nonumber \\
\label{SMRs}
& = & \frac{1}{4 \pi}  \left ( \frac{W}{\delta} \right )^4 \frac{ d\sigma/d{\Omega}}{\delta^2} \frac{ \Delta \omega}{\alpha(t)} \exp({ -{2L}/{l_s}})
\end{eqnarray}

This theoretical study is first applied to the experimental configuration described in the accompanying paper. The parameters used for the computation of the SMR are described in Tab.\ref{table}. 
The transport parameters of the paper ($l_s$, $l_t$, $R$) have been derived in the first section of this supplementary material. The experimental parameters ($\lambda$, $F$, NA, $W$, $\tau_c$) correspond to those reported in the paper.  The differential scattering cross-section $d \sigma/d\Omega$ has been estimated from Mie theory \cite{bohren1983absorption} by considering a 10 $\mu$m-diameter ZnO spherical bead as a target. The numerical value given in Tab.\ref{table} corresponds to an average of $d \sigma/d\Omega$ over the numerical aperture of the MO. At last, the Strehl ratio $S$ is estimated from  $l_c$, the coherence length of the ballistic wave-field in the focal plane, such that $S \sim (l_c/\delta)^2$. As $l_c \sim 5$ $\mu$m in the reported experiment and $\delta=\lambda/(2\mbox{NA})=1.6$ $\mu$m, it yields $S=0.1$.
 
\begin{figure}[h!]
\center
\includegraphics[width=8cm]{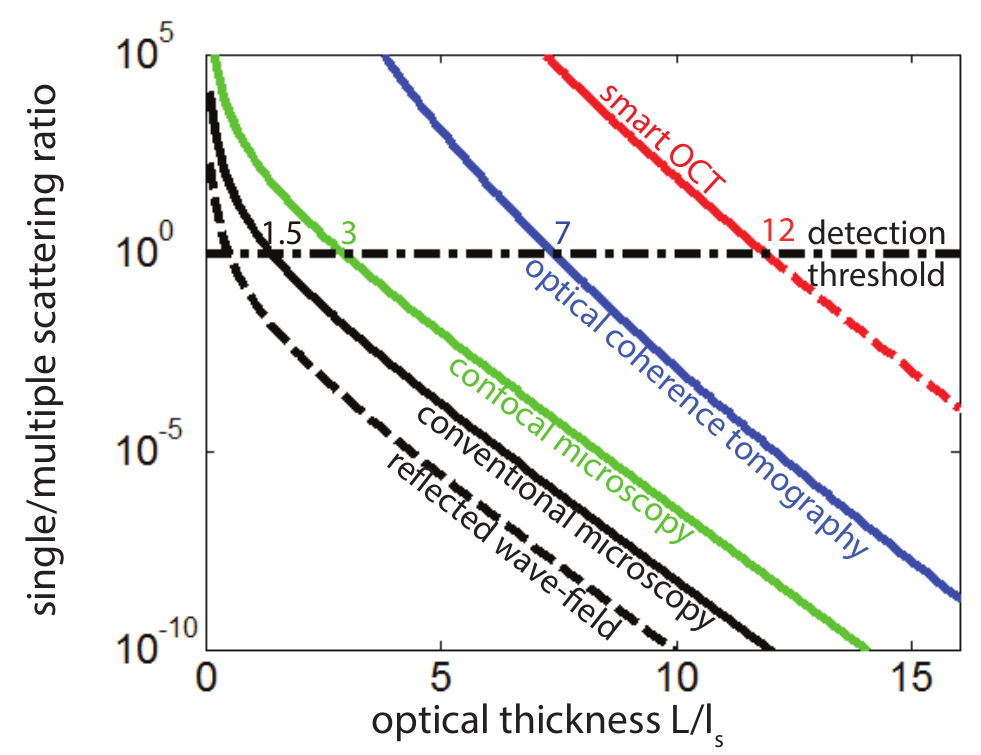}
\caption{\textbf{Theoretical prediction of the single-to-multiple scattering ratio in the experimental conditions of the article.} \label{fig0bis} Single-to-multiple scattering ratio for conventional microscopy [black line, Eq.\ref{SMRm}], confocal microscopy [green line, Eq.\ref{SMRc}], OCT [blue line, Eq.\ref{SMRt}] and smart OCT [red line, Eq.\ref{SMRs}] as a function of the optical thickness $L/l_s$. The y-axis is in log-scale. The detection threshold (SMR$\sim1$, black dashed horizontal line) yields an imaging depth limit of $\sim 2l_s$ for conventional microscopy, $\sim4l_s$ for confocal microscopy, $\sim 7 l_s$ for OCT and $\sim 12 l_s$ for smart OCT.}
\end{figure}
Figure~\ref{fig0bis} displays the evolution of the SMR as a function of the optical thickness $L/l_s$ by applying numerically Eqs.\ref{SMR}, \ref{SMRm}, \ref{SMRc}, \ref{SMRt} and \ref{SMRs} with the parameters shown in Tab.\ref{table}. Note that Eq.\ref{albedo_t} is considered for the computation of the time-dependent albedo $\alpha(t)$. Figure~\ref{fig0bis} illustrates how a confocal illumination and a coherence time gating allows to drastically improve the SMR compared to conventional microscopy. Nevertheless, the theoretical OCT imaging-depth limit (SMR$\sim 1$) remains limited to $7 l_s$ in this configuration. This explains why the time-gated confocal image is not able to reveal the presence of a target for an optical thickness $L\sim$ 12$l_s$ in our experiment. On the contrary, the smart-OCT imaging-depth limit is predicted to be around 12 $l_s$. This is in a remarkable agreement with our experimental results showing a successful target detection for an optical tickness $L\sim12.25$ $l_s$. Contrastedly, the target is far to be detectable and imaged whether it be by conventional microscopy (SMR$_m$ $\sim 10^{-10}$, Eq.\ref{SMRm}), confocal microscopy (SMR$_c$ $\sim 10^{-8}$, Eq.\ref{SMRc}) or by OCT (SMR$_t$ $\sim 10^{-5}$, Eq.\ref{SMRt}). The SMR of the backscattered wave-field is of $10^{-12}$ (Eq.\ref{SMR}) meaning that only 1 reflected photon over one thousand billions is associated to a single scattering event from the target in our experiment.

In Fig.~\ref{fig5}, the SMR is displayed as a function of the optical depth $F/l_s$ in the context of biological tissues imaging. Note that the SMR for conventional and confocal microscopy (Eqs.\ref{SMRm}-\ref{SMRc}) has been computed by considering the asymptotic limit of the static albedo ($\alpha \sim 3/(8\pi)$, see Eq.\ref{static_albedo}). As to coherence time gating, Eq.\ref{albedo_t2} is considered for the computation of the time-dependent albedo $\alpha(t)$ as the target is assumed to be embedded within the scattering medium. The parameters used for the computation of the SMR are described in Tab.\ref{table}. The considered transport parameters ($l_s$, $l_t$) are typical of in-vivo cortex tissues \cite{Schott}. The experimental parameters ($\lambda$, $F$, NA, $W$, $\tau_c$) are typical of full-field OCT \cite{Assayag}.  The internal reflection coefficient $R$ is assumed to be zero since the use of an immersion microscope objective will limit the impedance mismatch with tissues. The target scattering cross-section is arbitrarily chosen to be the same as in the reported experiment. At last, the Strehl ratio $S$ in brain tissues is estimated from a two photons microscopy experiment that reports a fivefold signal enhancement when optical aberrations from the brain tissues are corrected with adaptive optics \cite{Ji}.  
{\small
\begin{table}[h!]
\center
\begin{tabular}{c|c|c}
Experimental   & Reported experiment:  & Full-field oct:  \\
configuration  & paper sheets \cite{Hubbe} & Brain tissues \cite{Assayag}  \\
\hline
\hline
$n$ & 1.5 \cite{Hubbe}  & 1.4 \cite{Jacques} \\
$l_s$ [$\mu$m ] & 13.4 & 200 \cite{Schott} \\
$l_t$ [$\mu$m ] & 19.9 & 2000 \cite{Schott} \\
$R$ & 0.57 \cite{Xhu}  & 0 \\
$d\sigma/d\Omega$ [m$^2$.sr$^{-1}$]  & $8\times 10^{-11}$ & $8\times 10^{-11}$ \\
$S$ & 0.1  &  0.4 \\
$F$ [mm] & 1 & na \\
$\lambda$ [nm] & 810 & 810 \\
NA & 0.25 & 0.4 \\
$W$ [$\mu$m] & 40 & 1000 \\
$\tau_c$ [fs] & 50 & 5 \\
\hline
\end{tabular}
 \caption{\label{table}\textbf{Experimental parameters used for the theoretical prediction of the single-to-multiple scattering ratio in Figs.~\ref{fig5} and \ref{fig0bis} of the accompanying paper.}.}
\end{table}
}

\subsection{Eigenvalue distribution of the reflection matrix in the multiple scattering regime} 

\begin{figure*}[htbp]
\center
\includegraphics[width=17cm]{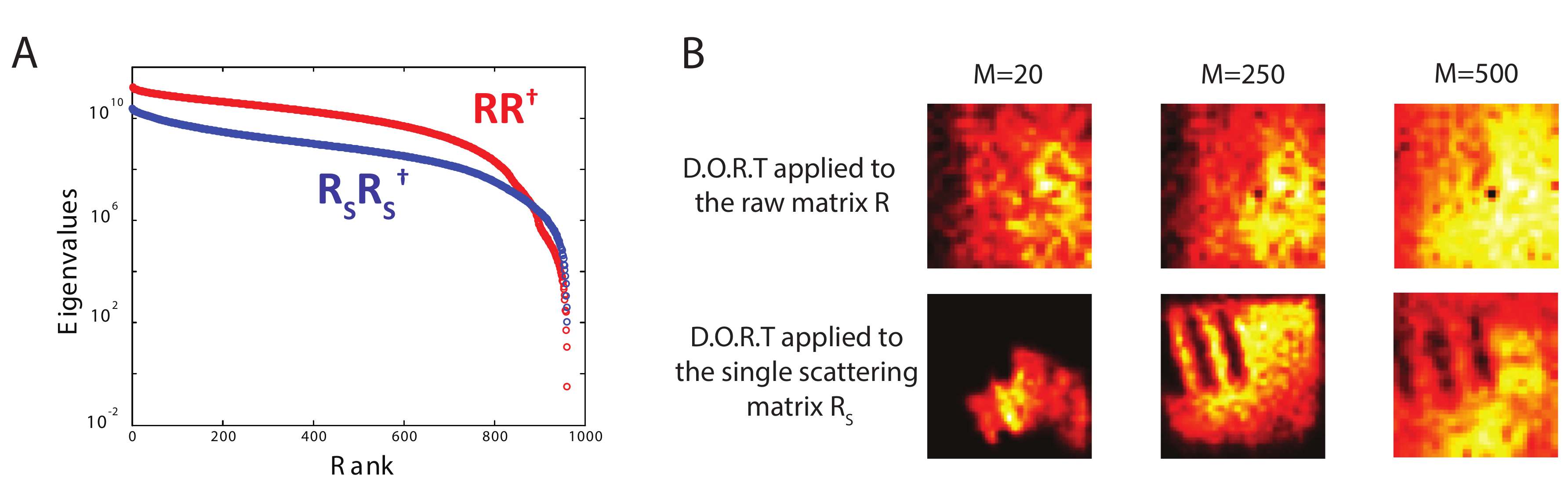}
\caption{\textbf{Iterative time-reversal processing applied to the raw reflection and single scattering matrices measured through biological tissues [Fig.\ref{fig4}].} (A)Eigenvalues $\sigma_i^2$ of the raw time reversal operator $\mathbf{R}\mathbf{R^{\dag}}$ (red line) and of the single scattering time reversal operator $\mathbf{R}_S\mathbf{R_S^{\dag}}$ (blue line). (B) Top and bottom lines: Images of the resolution target  by considering, from left to right, the 20, 250 and 500 first eigenstates of $\mathbf{R}$ and $\mathbf{R}_S$, respectively.}
\label{FigS3}
\end{figure*}
In this section, we derive the numerical method to obtain the eigenvalue distribution of $\mathbf{R_SR_S^{\dag}}$ expected in a fully multiple scattering regime [see Fig.~\ref{fig2}F]. If the coefficients of $\mathbf{R_S}$ were complex random variables independently and identically distributed, the eigenvalue distribution would follow the Marcenko-Pastur law \cite{marcenko,tulino}. However, this assumption is not fulfilled here. First, all the elements of $\mathbf{R_S}$ do not exhibit the same variance because of the single scattering filtering operation [see Figs.~\ref{fig2}E and \ref{fig3}E]. Second, some residual correlations may arise in the measured $\mathbf{R}-$matrix due to experimental imperfections. In particular, a slight curvature of the reference beam in the experiment may induce some short-range correlations of the wave-field at the output. 

The correlation between two coefficients $r_{il}$ and $r_{jm}$ of $\mathbf{R}$ can be expressed as \cite{sengupta}
\begin{equation}
\label{eqn:eq6}
\left < {r}_{il} {r}_{jm}^*\right>= \left < |{r}_{il}|^2\right> c_{ij}d_{lm}
\end{equation}
where the symbol $<.>$ denotes an ensemble average. $\mathbf{C}$ and $\mathbf{D}$ are $N \times N$ matrices. We will refer to them as the correlation matrices. As the correlation properties are statistically invariant by translation, $\mathbf{C}$ and $\mathbf{D}$ are Toeplitz matrices: $c_{ij}=C_{i-j}$ and $d_{lm}=D_{l-m}$.  The correlation coefficients, $C_n$ and $D_n$, between the columns and the lines of $\mathbf{R}$ can be estimated as follows
\begin{equation}
\label{eqn:corr_coeff}
C_n = \frac{\left < {r}_{i,j} {r}_{i,j+n}^*\right >_{(i,j)}}{\left < \left | {r}_{i,j}\right |^2 \right >_{(i,j)}} , \, \, D_n =  \frac{\left < {r}_{i,j}{r}_{i+n,j}^*\right >_{(i,j)}}{\left < \left | {r}_{i,j}\right |^2 \right >_{(i,j)}}
\end{equation}
where the symbol $<. >$ denotes an average over the variables in the subscript. From these correlation coefficients, one can estimate the correlation matrices $\mathbf{C}$ and $\mathbf{D}$.

Once $\mathbf{C}$ and $\mathbf{D}$ are estimated, one can deduce the eigenvalue distribution expected in a fully multiple scattering regime  \cite{aubry_wrmc}. It consists in generating numerically a matrix $\mathbf{P}$ whose elements are circularly symmetric complex gaussian random variables with zero mean. Then, a matrix $\mathbf{Q}$ is built from $\mathbf{P}$, such that
\begin{equation}
\label{eqn:eq8}
\mathbf{Q}=\mathbf{{C}}^{\frac{1}{2}}\mathbf{P}\mathbf{{C}}^{\frac{1}{2}}
\end{equation}
One can show that the matrix $\mathbf{Q}$ exhibits the same correlation properties at emission and reception as the experimental matrix $\mathbf{R}$. The $\mathbf{Q}$-matrix is then projected on the characteristic SS matrix $\mathbf{S}$ [Eq.3]
\begin{equation}
\label{eqn:eqS}
\mathbf{Q_S}=\mathbf{Q}\circ \mathbf{S}
\end{equation}
A singular value decomposition of $\mathbf{Q_S}$ is then performed. The eigenvalues of $\mathbf{Q_S}\mathbf{Q_S^{\dag}}$ (square of the singular values) are renormalized by their mean. An histogram of the eigenvalues is then obtained by averaging over 2000 {numerically generated random matrices $\mathbf{P}$}. The resulting theoretical distribution is shown in Fig.~\ref{fig2}F.


%

\end{document}